\documentstyle[epsfig,12pt,moriond]{article}
\def\mm{\,$\mu$m\,}

\begin{document}
\heading{Mid and Far Infrared Surveys with ISO}
\author{D. Elbaz $^{1}$, $^{2}$} {$^{1}$ CEA Saclay/DSM/DAPNIA/Service
	d'Astrophysique, Orme des Merisiers, F-91191 Gif-sur-Yvette
	Cedex, France} {$^{2}$ University of California, Santa Cruz
	(UCSC),Physics Department, Santa Cruz, CA 95064, USA}
\begin{moriondabstract}
We present the results of the extragalactic surveys performed at 15
and 175\mm using the ISOCAM and ISOPHOT instruments on-board the
Infrared Space Observatory (ISO). The number counts at both
wavelengths present a strong evolution. The preliminary study of the
individual galaxies responsible for this evolution suggest that more
star formation was hidden by dust at redshifts around $z=1-2$ and
imply that the population of luminous IR galaxies detected by IRAS in
the local universe must have played a major role in the formation and
evolution of galaxies and cannot be considered anymore as extreme and
rare cases.  
\end{moriondabstract}

\section{Introduction}
Several critical observational advances have taken place in the past
few years to elucidate the formation and evolution of galaxies and the
rates at which stars formed at different cosmological epochs. With the
advent of magnitude limited samples of spectroscopic redshifts, it
became possible to discuss general properties of galaxies through
their comoving luminosity density or star formation rate (SFR) as a
function of redshift (Lilly et al 1996, Madau et al 1996). Since then,
we became familiar with the study of a new astronomical character, the
'$1~Mpc^3$' box of the universe. At the same time, the plot of the
cosmic background was enriched by COBE (Puget et al 1996, Fixsen et al
1998, Lagache et al 1999, Hauser et al 1999, Dwek et al 1998) and the
HST (Pozzetti et al 1998), on the far infrared to sub-millimeter part
and on the optical-UV part respectively. These two global views of the
past history of galaxy formation and evolution gave a new life to the
debate that begun with the discovery by IRAS of Luminous InfraRed
Galaxies (LIRGs, see the review by Sanders \& Mirabel 1996). Galaxies
brighter than $10^{11}~L_{\odot}$ radiate most of their light in the
infrared (IR) range and exhibit the strongest star formation rates
ever measured. No other observations, else than direct infrared
observations, had prepared such a discovery. However, the infrared
radiation of nearby galaxies ($z<0.2$) makes only $30~\%$ of their
optical light and only $6~\%$ of it comes from LIRGs (based on IRAS
results, Soifer \& Neugebauer 1991), which consequently weight only
$2~\%$ by comparison with the optical light from 'normal'
galaxies. Prior to the COBE, SCUBA and ISO results, hence only three
years ago, the matter of debate could have been summarized by the
following question: should we consider LIRGs, as anything else than
extreme cases, non representative of the general behavior of galaxies
?

When the previously quoted authors, after Puget et al (1996), found
from the COBE FIRAS and afterwards DIRBE measurements, that the
energetic content of the IR to sub-mm cosmic background was as high as
that of the UV to near-IR background, the debate became more intense
but still left opened the possibility to keep a scenario consistent
with the one deduced from the '$1~Mpc^3$' box of the universe story,
i.e. that one could reproduce both the CIRB and the variation of the
comoving UV luminosity density and SFR with redshift without the
recourse of LIRGs (see Fall, Charlot \& Pei 1996).

Several observations have now ruled out this picture. The redshift
dependance of the comoving star formation rate of the universe needs
not only the UV luminosity density to be known as a function of
redshift but also the IR luminosity density, which is dominated by
LIRGs at $z\sim1$ and above (see below). The most directs of these
observations, because they measure the IR light radiated by dust, come
from three instruments: ISOCAM (Cesarsky et al 1996), in the mid-IR,
and ISOPHOT (Lemke et al 1996), in the far-IR, on board ISO (Kessler
et al 1996), and the SCUBA bolometer array at JCMT, in the sub-mm (see
Lilly et al, in these proceedings). All three instruments present
number counts much above expectations from no evolution models based
on IRAS results on the local universe, indicating that the bulk of the
cosmic IR background (CIRB) originates from LIRGs, or even brighter
galaxies. Hence, answering NO to the previous question. So that we can
say that we are leaving in the middle of an ``Infrared
Revolution''. The following discussion will illustrate the role that
the ISOCAM and ISOPHOT extragalactic results are playing in it, being
complementary to and consistent with the SCUBA results.

\section{Description of the Surveys and Number Counts}
ISOCAM is a mid-IR camera which allows to perform either wide band
imagery or low resolution spectroscopy using Circular Variable
Filters, in the range 5-18\,$\mu$m. The two wide band filters that
were used for the extragalactic surveys, LW2 \& LW3, are centered at
respectively 6.75\,$\mu$m (5-8\,$\mu$m) and 15\,$\mu$m (12-18\,$\mu$m)
and were chosen because they cover the mid-IR emission from two
different origins, the aromatic features and the hot dust continuum
(see Section~\ref{origin}). Its pixel size is 6'' in the most
sensitive mode, for a FWHM PSF of 9'' at 15\,$\mu$m, but its
astrometric accuracy can reach 2'' in the micro-scanning mode of the
deepest surveys.
{\scriptsize
\begin{table*}[!ht]
\begin{center}
\leavevmode
\vspace{0.5 em}
\caption{\em Table of all ISOCAM extragalactic surveys sorted by
increasing depth. The results used for the number counts are denoted
by a $\star$. Col.(4) is the integration time per sky position. The
depth of each survey is given in Col.(5), where the number between
parenthesis correspond to the 80 $\%$ completeness limit, and the
asscociated numbers of sources are given under Col. \#7 and \#15 for
the 7 and 15\,$\mu$m bands respectively. The coordinates are in
J2000. References: a-in preparation, 1-Siebenmorgen et al (1996),
2-Rowan-Robinson et al (1999), 3-Clements et al (1999), 4-Flores et al
(1999a,b), 5-Metcalfe et al (1999), 6-Aussel et al (1999), 7-Taniguchi
et al (1997a), 8-Taniguchi et al (1997b), 9-Altieri et al (1999).}
\begin{tabular}{|l|ll|ll|ll|ll|ll|llll|}
\hline
Field Name      &$\lambda$&  & Area&       & Int.& & depth& &\#7 &\#15 &h & m & $^{\circ}$ & ' \\
              	&$\mu m$& & ('$^2$)&     & (min)& & (mJy)& &     & &  & & & \\
\hline
CAM parallel$^1$
& 7&15		& 1.2e5&1.2e5	& $>$5&$>$5	& $>$2.5 &$>$2.5 & $>$1e4&$>$1e4&- & & & \\
ELAIS$^2$
& 7&15		& 2e4&4e4	& 0.7&0.7	& 1&3		& 1104&1618	&- & & & \\
LH Shallow$^{\star,a}$
& -&15		& -&1944	& 3&-		& -&0.72 (1)	& -&180 (80)	&10& 52.08 & +57&21.07 \\ 
Comet Field$^3$
& -&12		& -&360 	& 10&-		&- &0.5		& -&37		&03& 05.50 & $-$09&35.00\\
CFRS 14+52$^4$								    	       	      
& 7&15		& 100&100     	& 18&11		& 0.3&0.4	& 23&41		&14& 17.90 & +52&30.52 \\
LH Deep$^{\star,a}$
& 7&15		& 510&510	& 18&11		& 0.3&0.4 (0.6)	& -&166 (70)	&10& 52.08 & +57&21.07 \\
CFRS 03+00$^a$									    	       	      
& 7&15		& 100&100     	& 6&22		& 0.5&0.3	& -&-		&03& 02.67 & +00&10.35 \\
MFB Deep$^{\star,a}$
& 7&15		& 900&900	& 15.4&15.4	& 0.19&0.32 (0.4)&-&180 (144)	&03& 13.17 & $-$55&03.82 \\
A370$^5$										    	       	      
& 7&15		& 31.3&31.3    	& 42&42 	& -&0.26	& -&18		&02& 39.83 & $-$01&36.75\\
Marano UD$^{\star,a}$								    	       	      
& 7&15		& 85&90		& 120&114	& 0.18&0.14 (0.2)&-&142 (82)	&03& 14.73 & $-$55&19.58\\ 
MFB UD$^{\star,a}$
& 7&15		& 89&90		& 120&114	& 0.08&0.14 (0.2)& 115&137 (100)&03& 13.17 & $-$55&03.82 \\
A2218$^5$										    	       	      
& 7&15		& 16&16       	& 84&84 	& -&0.12 	& -&23		&16& 35.90 & +66&13.00 \\
HDF North$^{\star,6}$
& 7&15		& 10&24       	& 116&135  	& 0.05&0.1	& 7&44		&12& 36.82 & +62&12.97 \\
HDF South$^{\star,a}$								    	       	      
& 7&15		& 28&28       	& -&168  	& 0.05&0.1	& -&63		&22& 32.92 & +60&33.30 \\
Deep SSA13$^7$									    	       	      
& 7&-		& 9&-          	& 567&-      	& -&-		& -& -		&13& 12.43 & +42&44.40 \\
LH PGPQ$^8$								    	       	      
& 7&-		& 9&-          	& 744&-      	& 0.034&	& 15&-		&10& 33.92 & +57&46.30 \\
A2390$^{\star,9}$
& 7&15		& 5.3&5.3     	& 432&432 	& 0.026&0.054	& 32&31		&21& 53.57 & +17&40.18 \\ 
\hline        									   
\end{tabular}
\label{TABLE:cam}
\end{center}
\end{table*}
}

ISOPHOT is an imaging photo-polarimeter covering the 2.5-240\,$\mu$m
range. The most performant filter for extragalactic surveys is
centered at 175\,$\mu$m (130-219\,$\mu$m) and uses the four pixels
camera, C200, with a 92'' side length and a 1.9' FWHM PSF at
175\,$\mu$m.

{\scriptsize
\begin{table*}[!ht]
\begin{center}
\leavevmode
\vspace{0.5 em}
\caption{\em Table of ISOPHOT-FIRBACK extragalactic
surveys. References: 1-Oliver et al, in preparation, 2-Dole et al
1999, Puget et al 1999.}
\begin{tabular}{|l|ll|ll|ll|llllll|}
\hline
Field Name      &$\lambda$	&  	& Area    &     & depth	& 	&h & m & s & $^{\circ}$ & ' & "\\
              	&$\mu m$  	&  	&(sq.deg.)&     & (mJy)	& 	& &  &  &  &  & \\
\hline
ELAIS N1$^{1,2}$
& 90	&175	& 2.6	&3	& 0.1 & 0.1&16&10 &01 &$+$54 & 30& 36\\
ELAIS N2$^{2}$
& 90&175	& 2.7      &3	& 0.1 & 0.1&16&36 &58 &$+$41 & 15& 43 \\
MF FIRBACK$^2$
& -&175	        & -      &1	& - & 0.1&03 &11 & 00 &$-$54 & 45& 00 \\
\hline        									 \end{tabular}
\label{TABLE:phot}
\end{center}
\end{table*}
} 
The list of all ISOCAM extragalactic surveys performed during the ISO
lifetime is given in the Table~\ref{TABLE:cam}. We used 8 of them to
produce the number counts in the Fig.~\ref{FIG:integ} (from Elbaz et
al 1999), for a total of 614 sources above an 80 $\%$ completeness
limit. The ISOPHOT number counts from Dole et al (1999) superimposed
in the Fig.~\ref{FIG:integ} (left) use 208 sources brighter than 100
mJy in $\sim$3 square degrees from the fields quoted in the
Table~\ref{TABLE:phot}. The ISOCAM and ISOPHOT surveys were performed
in regions selected for their low Galactic foreground emission, both
in the northern and southern hemispheres to avoid strong contamination
from large-scale structure.
\section{Origin of the infrared emission}
\label{origin}
The origin of the mid-IR emission of galaxies has been a subject of
debate prior to IRAS, when evolved stars such as OH/IR stars were
expected to play a major role. However, based on the IRAS Point Source
Catalog, source counts at 12\,$\mu$m have shown that circumstellar
dust shells accounted for only $\sim$10$\%$ of the 12\,$\mu$m emission
in the Galaxy at $\mid b \mid \geq 10^{\circ}$ (see Review by Cox \&
Mezger 1989).  It is now established that except for ellipticals (with
no dust lanes), the mid-IR emission of galaxies does not originate
primarily from individual stars but rather from the light of young
stars (or of an AGN) absorbed by dust and re-radiated in the mid-IR
mostly at the interface between HII regions and molecular clouds (the
photo-dissociation regions, PDR, see Cesarsky \& Sauvage 1999 and
references therein). The mid-IR spectrum of a galaxy can be divided
into three components: a set of broadband aromatic features, with
their underlying continuum, narrow forbidden lines of ionized gas
(negligible in ISOCAM wide band imagery), and a hot thermal continuum
due to very small grains (VSGs) of dust heated by stellar radiation at
temperatures greated than 150\,K. The broadband aromatic features have
been proposed to originate either from PAHs (for Polycyclic Aromatic
Hydrocarbons, L\'eger $\&$ Puget 1984, Allamandola et al 1989) or from
coal grains (Papoular 1999). In either case, they require the presence
of aromatic structures, but their exact nature has not yet been
definitively established. The low resolution spectra from ISOCAM have
clearly established that in the wavelength range 5-8.5\,$\mu$m of the
7\,$\mu$m filter, the spectra of spiral and starburst galaxies were
dominated by the aromatic broadband features located at 6.2, 7.7, 8.6,
11.3 and 12.7\,$\mu$m. At redshifts of the order of z$\sim$0.8, the
median redshift of the faint ISOCAM sources (see
Section~\ref{natcam}), the 15\,$\mu$m band is shifted to this region
of the spectrum dominated by the aromatic features, with a
contribution of hot dust continuum.

\begin{figure}[!b]
\centerline{\psfig{file=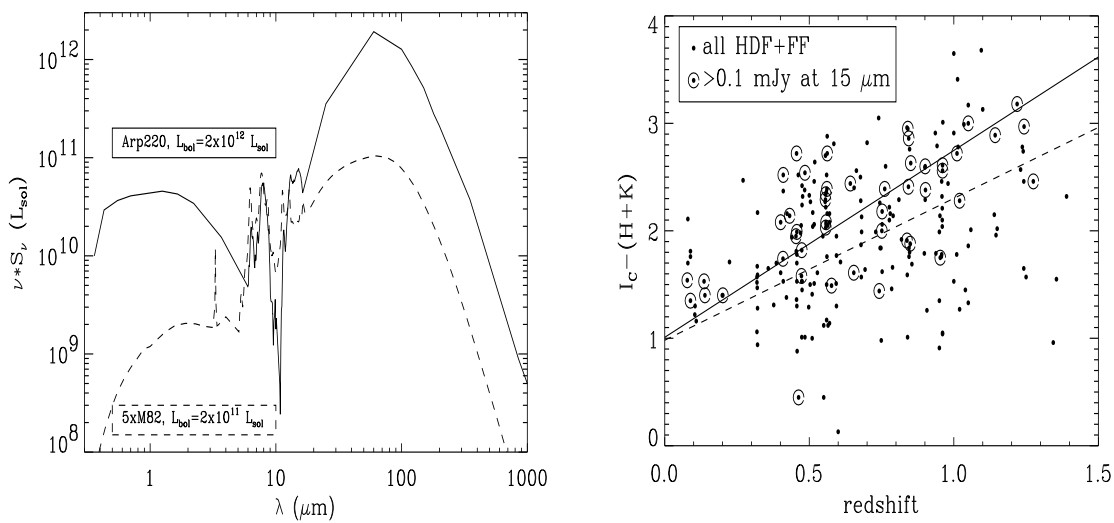}}
\caption{\em (left:) Comparison of the Arp 220 and M82 SEDs. (right:)
I(Kron-Cousins)-(H+K) color (from Barger et al 1999) versus redshift
for field galaxies (dots) and galaxies detected above 0.1 mJy at
15\,$\mu$m by ISOCAM (dots surrounded by circles), in the Hubble Deep
Field North plus its Flanking Fields. Plain line: linear fit to the
colors of ISOCAM galaxies. Dashed line: idem for all galaxies.}
\label{FIG:comp_hdf}
\end{figure}

At larger wavelengths, the spectral energy distribution (SED) of a
galaxy is progressively dominated by the thermal emission of larger
grains heated at a lower temperature (around 60\,K), which produces
the bulk of the overall IR emission originating from dust (around
60-100\,$\mu$m). This 'cold' dust dominates the IR emission measured
by ISOPHOT. At 175\,$\mu$m, the physical origin of the emission
remains the same up to high redshifts and is well-known to be
correlated with star formation and radio activity (Helou, Soifer \&
Rowan-Robinson 1985).

The link between mid-IR and star formation (SF) is less
straightforward, but only LIRGs can be detected at $z\sim0.8$ above
0.1 mJy at 15\,$\mu$m. Hence mid-IR surveys can at least pinpoint IR
active galaxies. Moreover, a correlation was found between mid-IR and
UV flux densities (Boselli et al 1997, 1998) as well as with
$H_{\alpha}$ (for the disks of spiral galaxies, Roussel et al, these
proceedings). However, the complex link between mid-IR and SF is
illustrated by the Fig.~\ref{FIG:comp_hdf} (left), where we compare
the SED of the ultra-LIRG Arp 220 and the nearby starburst M82,
normalized by a factor 5 to reach the same energy density at $\sim$8
\,$\mu$m.  At z$\sim$0.8, both galaxies would have the same 15\,$\mu$m
flux density but differ by a factor $\sim$20 in the far-IR. But Arp220
is an extreme case which can easily be rejected as a candidate SED for
the ISOCAM galaxies (see Section~\ref{natcam}), and M82's SED, which
is more typical of most LIRGs ($L_{bol}\sim L_{IR}>10^{11}~L_{\odot}$)
and ultra-LIRGs ($L_{bol}>10^{12}~L_{\odot}$) although it is only
$4~10^{10}~L_{\odot}$, is probably a better candidate. Indeed, in the
HDF-N seven galaxies, which were detected both at 15\,$\mu$m and in
the radio (Richards et al 1998), show consistent SF rates if we use
the SED of M82 to convert the mid-IR flux into a SF rate (Aussel et al
1999, in preparation).
\section{Discussion about the 15\,$\mu$m results}
\label{isocam}
\subsection{The 15\,$\mu$m counts}
\begin{figure}[!b]
\centerline{\psfig{file=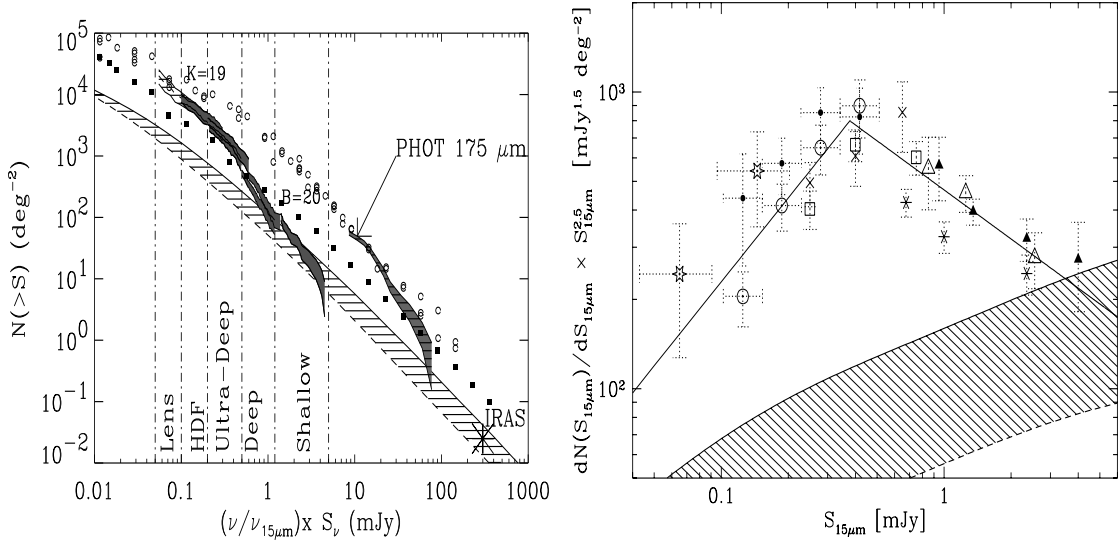}}
\caption{\em (Left:) Integral counts, i.e. the number of galaxies, N,
detected at 15 and 175\,$\mu$m (from Dole et al 1999, normalized in
$\nu S_{\nu}$) above the flux S(mJy), with 68~\% confidence contours.
K counts (Gardner et al 1993) and B counts (Metcalfe et al 1995),
multiplied by the ratio $\nu/\nu_{15}$ to represent the relative
energy densities, are superimposed with open circles and filled
squares, respectively.  The hatched area materializes the range of
possible expectations from models assuming no evolution at 15\,$\mu$m
and normalized to the IRAS 12\,$\mu$m local luminosity function
(LLF). (Right:) Differential Number Counts of the 15\,$\mu$m Galaxies,
with 68\% error bars. The counts are normalized to a Euclidean
distribution ($\alpha=-2.5$).}
\label{FIG:integ}
\end{figure}

The first striking result of the 15\,$\mu$m source counts is the
consistency of the eight surveys (noted with a $\star$ in the
Table~\ref{TABLE:cam}) over the full flux range
(Fig.~\ref{FIG:integ}). Some scatter is nevertheless apparent; given
the small size of the fields surveyed, we attribute it to clustering
effects.  The two main features of the observed counts are a
significantly super-euclidean slope ($\alpha=-3.0$) from 3 to 0.4 mJy
and a fast convergence at flux densities fainter than 0.4 mJy. In
particular, the combination of five independent surveys in the flux
range 90-400\,$\mu$Jy shows a turnover of the normalized differential
counts around 400\,$\mu$Jy and a decrease by a factor $\sim 3$ at
100\,$\mu$Jy.  We believe that this decrease, or the flattening of the
integral counts, below $\sim$400\,$\mu$Jy, is real.  It cannot be due
to incompleteness, which was quantified using the Monte-Carlo
simulations.  The differential counts can be fitted by two power laws
by splitting the flux axis in two regions around 0.4 mJy. In units of
mJy$^{-1}$ deg$^{-2}$, we obtain, by taking into account the error
bars ($S$ is in mJy):
\begin{eqnarray}  
\frac{dN(S)}{dS} = \left\{ 
\begin{array}{cccc}
(2000\pm600)& S^{(-1.6\pm0.2)} &\ldots&  0.1\le S\le0.4\\
&&&\\
(470\pm30)& S^{(-3.0\pm0.1)} &\ldots& 0.4\le S\le4\\
\end{array}
\right.
\end{eqnarray} 

The total number density of sources detected by ISOCAM at 15\,$\mu$m
down to 100\,$\mu$Jy (no lensing) is ($2.4\pm0.4$) arcmin$^{-2}$. It
extends up to ($5.6\pm1.5$) arcmin$^{-2}$, down to 50\,$\mu$Jy, when
including the lensed field of A2390 (Altieri et al 1999).  The
differential counts (Fig.~\ref{FIG:integ} (right)), which are
normalized to $S^{-2.5}$ (the expected differential counts in a non
expanding Euclidean universe with sources that shine with constant
luminosity), present a turnover around $S_{15\,\mu m}$=0.4 mJy, above
which the slope is very steep ($\alpha=-3.0\pm0.1$).  No evolution
predictions were derived assuming a pure {\it k}-correction in a flat
universe ($q_0=0.5$), including the effect of the aromatic features in
the galaxy spectra. The lower curve of the hatched area, which
materializes the 'no evolution' region in the Fig.~\ref{FIG:integ}, is
based on the Fang et al (1998) IRAS 12\,$\mu$m local luminosity
function (LLF), using the spectral template of a relatively quiescent
spiral galaxy (M51). The upper curve is based on the Rush, Malkan \&
Spinoglio (1993) IRAS-12\,$\mu$m LLF, translated to 15\,$\mu$m using
as template the spectrum of M82, which is typical of most starburst
galaxies in this band. More active and extincted galaxies, like
Arp220, would lead to even lower number counts at low fluxes while
flatter spectra like those of AGNs are less flat than the one of
M51. In the absence of a well established LLF at 15\,$\mu$m, we
consider these two models as upper and lower bounds to the actual
no-evolution expectations; note that the corresponding slope is $\sim
-2$. The actual number counts are well above these predictions; in the
0.3\,mJy to 0.6\,mJy range, the excess is around a factor 10: clearly,
strong evolution is required to explain this result (note the analogy
with the radio source counts, Windhorst et al 1993).
 
For comparison, we have superimposed on Fig.~\ref{FIG:integ} the
optical B band (Metcalfe et al 1995) and near-IR K band (Gardner et al
1993) integral counts, normalized in terms of $\nu S_{\nu}$.  For
bright sources, with densities lower than 10 deg$^{-2}$, these curves
run parallel to an interpolation between the ISOCAM counts presented
here and the IRAS counts; the bright K sources emit about ten times
more energy in this band than a comparable number of bright ISOCAM
sources at 15\,$\mu$m. But the ISOCAM integral counts present a rapid
change of slope around 1-2 mJy, and their numbers rise much faster
than those of the K and B sources. The sources detected by ISOCAM are
a sub-class of the K and B sources which harbor activity hidden by
dust.
\subsection{Nature of the 15\,$\mu$m galaxies} 
\label{natcam} 
\begin{figure}
\centerline{\psfig{file=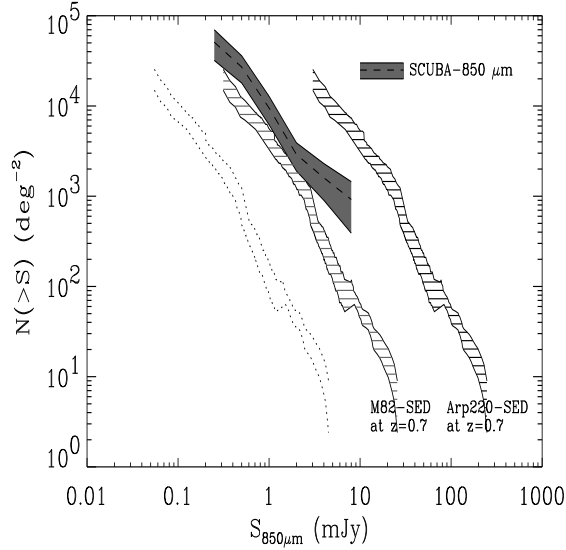,height=8.0cm,width=8.0cm}}
\caption{\em Integral counts at 850\,$\mu$m from SCUBA (Blain et
al 1999). From the 15\,$\mu$m ISOCAM counts (dotted lines), one can
extrapolate the contribution of the ISOCAM galaxies at 850\,$\mu$m for
a given redshift distribution and spectral energy distribution
(SED). Two examples are shown here using the SEDs of M82 and Arp 220,
and assuming that all ISOCAM galaxies have the same redshift, z=0.7,
corresponding to their median redshift in the HDF and CFRS
fields.}
\label{FIG:scuba}
\end{figure}

We believe, according to the results obtained on the HDF (Aussel et al
1999a,b) and CFRS fields (Flores et al 1999a,b, see also these
proceedings), that the sources responsible for the 'bump' in the
15\,$\mu$m counts are not the low mass faint blue galaxies which
dominate optical counts and have a median redshift around z$\sim$0.6
(Pozzetti et al 1998).  Instead, they most probably are bright and
massive galaxies whose emission occurs essentially in the IR and could
account for a considerable part of the star formation in the universe
at $z<1.5$. Indeed, from the full sample of 15\,$\mu$m galaxies with
known redshift and optical-near IR magnitudes, we found that these
galaxies are massive ($\sim 10^{11}~M_{\odot}$) and that their
emission occurs essentially in the IR. Their median redshift is
z$\sim$0.8 in a sample of 42 galaxies brighter than 0.1 mJy in the
HDF+FF (Aussel et al 1999b) and z$\sim$0.7 in a sample of 41 galaxies
brighter than 0.35 mJy in the CFRS-1415 field (Flores et al
1999b). The assessment of their bolometric luminosity requires the
assumption of a spectral energy distribution, which is largely
uncertain since the ratio of the far-IR over mid-IR flux densities is
highly variable among galaxies. However, one can set limits to the
bolometric luminosity using the following arguments. If all galaxies
had extreme SEDs like the one of Arp220, they would produce a
contribution to the SCUBA-850\,$\mu$m number counts (see Blain et al
1999) much above observations (see Fig.~\ref{FIG:scuba}). An
Arp220-like SED can also be rejected for over-producing the
140\,$\mu$m EBL with respect to the one measured by DIRBE on-board
COBE (see Fig.~\ref{FIG:cirb}). The light radiated in the far-IR by
these galaxies however cannot be much lower than that emitted at
15\,$\mu$m, since only galaxies luminous in the IR can be detected at
z$>$0.5 with a 15\,$\mu$m flux density larger than 0.1 mJy (the origin
of the emission cannot be stellar without requiring excessive masses).
We therefore estimate the mean bolometric luminosity of these galaxies
to be of the order of a few $10^{11}~L_{\odot}$.

The galaxies from the HDF-N plus Flanking Fields whose 15\,$\mu$m flux
density is greater than 0.1 mJy (sensitivity limit of ISOCAM) harbour
a I$_C$-(H+K) color distribution very similar to that of the whole
sample of galaxies for which we have access to both the I$_C$-(H+K)
colors and spectroscopic redshifts (see Fig.~\ref{FIG:comp_hdf}
(right)). Aussel et al (1999b and PhD thesis) showed that the
sub-sample of galaxies with known spectroscopic redshift keeps the
same color properties than the full sample of HDF+FF galaxies, hence
we did not include a selection bias in the color distribution.

The median colors are I$_{C}$-(H+K)= 2.3 for the 15\,$\mu$m galaxies
and I$_{C}$-(H+K)= 2 for all galaxies, hence only differ by 0.3
dex. However, the linear fit to the I$_C$-(H+K) versus redshift plot
of the two samples of galaxies in Fig.~\ref{FIG:comp_hdf} (right)
shows that the 'dusty' galaxies tend to redden slightly faster with
increasing redshift, than the natural reddening of the whole
population of field galaxies which is due to {\it k}-correction: the
color difference increases from 0.2 dex below z=0.7 to 0.4 dex above
z=0.7. Nevertheless, this difference is not strong enough to allow one
to separate the infrared galaxies from the whole sample like for
Lyman-break galaxies (even accounting for the other set of optical
colors existing for the HDF galaxies). The origin of this reddening
with redshift is not clear but it is probably due to a selection of
the galaxies suffering from more extinction, hence emitting more in
the infrared, at larger redshifts.

Hence the population of galaxies producing the 15\,$\mu$m number
counts excess is very distinct from the one which dominates the deep
optical counts, known to be made of low mass galaxies with blue
luminosities, although at similar redshifts. In other words, the star
formation activity responsible for the light emitted by the 15\,$\mu$m
galaxies is not the other face of the same star formation activity
already quantified from the optical surveys, but instead it should be
considered as a second component, which was previously missed.
\section{Discussion about the 175\,$\mu$m results}
\label{isophot}
\subsection{The 175\,$\mu$m counts}
\begin{figure}
\centerline{\psfig{file=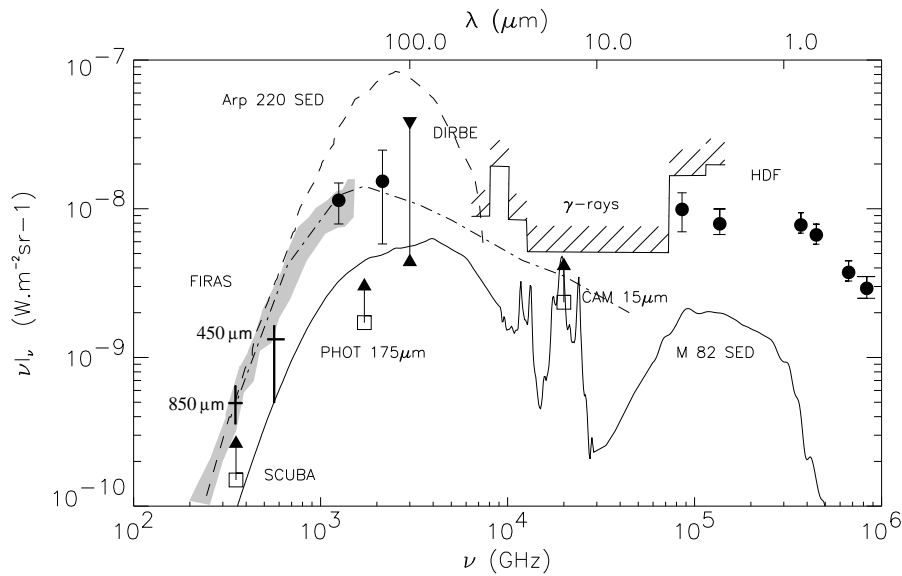,height=10.0cm,width=14.0cm}}
\caption{\em Cosmic background from UV to mm. Open squares give
the lower limits from ISO (ISOCAM-15\,$\mu$m and ISOPHOT-175\,$\mu$m)
and SCUBA-850\,$\mu$m (the two thick crosses correspond to the recent
values at 450 and 850\,$\mu$m from Blain et al 1999). The optical-UV
points are from Pozzetti et al (1998). The COBE FIRAS (grey area) and
DIRBE 140 \& 240\,$\mu$m (filled circles) data are from Lagache et al
(1999).}
\label{FIG:cirb}
\end{figure}

As for the 15\,$\mu$m ISOCAM counts, the 175\,$\mu$m ISOPHOT counts
present a strong excess with respect to IRAS of about one order of
magnitude (Puget et al 1999, Dole et al 1999). The slope of the
integral counts is $\alpha=-2.2$, hence much above the Euclidean value
of -1.5, and far from flattening, contrary to the ISOCAM counts which
go deep enough to detect the beginning of the flattening, with a slope
decreasing from -2.4 around 0.5 mJy to -1.2 around 0.1 mJy. It is
therefore natural that the detected sources account only for 10 $\%$
of the CIRB measured by COBE (see Fig.~\ref{FIG:cirb}).

\subsection{Nature of the 175\,$\mu$m galaxies} 
\label{natphot} 
The search for optical counterparts to the ISOPHOT-175\,$\mu$m
galaxies is not an easy task because of the large uncertainty on the
position of the ISOPHOT sources and of the absence of a clear optical
signature. Hence the use of other wavelengths with better spatial
resolution and sensitivity to the IR activity is mandatory in this
study. Because of the good radio/far-IR correlation, the radio domain
is an obvious candidate and VLA surveys of the northern fields (ELAIS
N1 \& N2) have been carried out by Ciliegi et al (1999) in the
framework of the ELAIS project (Rowan-Robinson et al 1999). A sample
of ten 21 cm radio counterparts of the ISOPHOT galaxies, i.e. with
$S_{175\,\mu m}>~100~mJy$, have been selected for their high $175\,\mu
m/21~cm$ ratio by Scott et al (1999) for a follow-up at 450 and
850\,$\mu$m with SCUBA. Assuming a dust temperature of $T_d=40~K$,
they find a range of low to moderate redshifts of $z=0-1.5$ (no
spectroscopic redshift are available yet), when accounting for the
SCUBA measurements or upper limits together with the ISOPHOT ones.

In these two fields, the ISOCAM surveys are not sensitive enough to
strongly constrain the SED of the ISOPHOT galaxies and miss the
majority of them ($S_{15\,\mu m}>~3~mJy$). In the southern field
however (called MFB, for 'Marano FIRBACK' since it is shifted with
respect to the original Marano Field, Marano, Zamorani \& Zitelli,
1988, in order to minimize the foreground Galactic cirrus emission),
deep ISOCAM surveys have been performed over 0.5 square degree down to
0.3 mJy (and a completeness limit of $0.4~mJy$). Half of the 24
FIRBACK galaxies in the field have a counterpart, sometimes several,
at 15\,$\mu$m above $S_{15\,\mu m}=~0.3~mJy$ and closer than 30''
(except one at 50''). One can compare their $S_{175\,\mu m}/S_{15\,\mu
m}$ ratio to the one of any template SED at any redshift and determine
a possible redshift range for this SED. The striking result of this
study is that none of the 12 galaxies present any 15\,$\mu$m
counterpart compatible with the SED of M82 at any redshift (unless
they are brighter than $10^{13}~L_{\odot}$), while all but one of them
are consistent with an Arp220-like galaxy of similar luminosity
($\sim~10^{12}~L_{\odot}$) and at a redshift between z=0.1-0.4. For
the remaining 12 galaxies, they could be compatible with an M82-like
SED for $z>1.2$ and $L_{bol}>4~10^{12}~L_{\odot}$ or with an
Arp220-like SED (though fainter), if only their redshift is larger
than $z=0.3$. In any case, the FIRBACK galaxies are most probably
LIRGs or ULIRGs and their large number density implied by the number
counts implies a strong evolution of this population of galaxies with
respect to IRAS, consistently with ISOCAM, although the 175\,$\mu$m
galaxies are most probably brighter than the 15\,$\mu$m ones.

\section{Conclusion}
The ISO extragalactic surveys have demonstrated the fundamental role
of IR in the understanding of galaxy evolution. However, only
ground-based follow-ups will allow us to understand the nature of the
population of galaxies responsible for the strong evolution exhibited
by the IR number counts. Are they particularly rich in metals ? Do
they harbor active galactic nuclei ? We have a large enough sample of
galaxies to be able to quantify the evolution of the luminosity
function in the IR as a function of redshift. However, we are still
far from understanding the origin of the huge cosmic IR background
revealed by COBE, but SIRTF and FIRST are still to come...


%
\vfill

\begin{moriondbib}
\bibitem{} Allamandola, L.J., Tielens, A.G.G.M., Barker, J.R., 1989, ApJS 71, 733
\bibitem{} Altieri, B., Metcalfe, L., Kneib, J.P., 1999, A\&A  343, L65
\bibitem{} Aussel, H., Cesarsky, C., Elbaz, D., Starck, J.L., 1999a, A\&A 342, 313
\bibitem{} Aussel H., Elbaz D., D\'esert F.X. et al, 1999b, in Cox,
P., Kessler, M.F. (eds.), {\em The Universe as seen by ISO}, ESA
SP-427, p. 1023
\bibitem{} Barger, A.J., Cowie, L.L., Trentham, N., et al, 1999, AJ 117, 102
\bibitem{} Blain et al, 1999, to appear in `The Hy-redshift universe',
eds A. Bunker \& W. van Breughel, astro-ph/9908111
\bibitem{} Boselli, A., Lequeux, J., Contoursi, A., et al, 1997, A\&A 324, L13
\bibitem{} Boselli, A., Lequeux, J., Sauvage, M., et al, 1998, A\&A 335, 53
\bibitem{} Cesarsky, C., Abergel, A., Agn\`ese, P., et al, 1996, A\&A 315, L32
\bibitem{} Cesarsky, C., Sauvage, M., 1999, 'Millenium Conference on
Galaxy Morphology' (D. Block, I. Puerari, A. Stockton, dW. Ferreira
Eds), Johannesburg, South Africa, astro-ph/9909369
\bibitem{} Ciliegi, P., MacMahon, R.G., Miley, G., et al, 1999, MNRAS 302, 222
\bibitem{} Clements, D.L., D\'esert, F.X., Franceschini, A., et al, 1999, A\&A 346, 383
\bibitem{} Cox, P., Mezger, P.G., 1989, A\&AR 1, 49
\bibitem{} Elbaz, D., Cesarsky, C.J., Fadda, D., et al, 1999, A\&AL, in press, astro-ph/9910406
\bibitem{} Fall, S., Charlot, S., Pei, Y.C., 1996, ApJ, 464, L43
\bibitem{} Fang, F., Shupe, D., Xu, C., Hacking, P., 1998, ApJ 500, 693
\bibitem{} Fixsen, D.J., Dwek, E., Mather, J.C., et al, 1998, ApJ 508, 123
\bibitem{} Flores, H., Hammer, F., D\'esert, F.X., et al, 1999a, A\&A 343, 389
\bibitem{} Flores, H., Hammer, F., Thuan, T., et al, 1999b, ApJ 517, 148
\bibitem{} Gardner, J.P., Cowie, L.L., Wainscoat, R.J., 1993, ApJ 415, L9
\bibitem{} Hauser, M.G., Arendt, R.G., Kelsall, T., 1998, ApJ 508, 25
\bibitem{} Helou, G., Soifer, B.T., Rowan-Robinson, M., 1985, ApJ 298, L7
\bibitem{} Kessler, M., Steinz, J., Anderegg, M., et al, 1996, A\&A 315, L27
\bibitem{} Lagache, G., Abergel, A., Boulanger, F., et al, 1999, A\&A 344, 322
\bibitem{} L\'eger, A., Puget, J.L., 1984, A\&A 137, L5
\bibitem{} Lemke, D., Klaas, U., Abolins, J., et al, 1996, A\&A 315, L64
\bibitem{} Lilly, S., Le F\`evre, O., Hammer, F., et al, 1996, ApJ 460, L1
\bibitem{} Madau, P., Ferguson, H., Dickinson, M., et al, 1996, MNRAS 283, 1388
\bibitem{} Marano, B., Zamorani, G., Zitelli, V., 1988, MNRAS 232, 111
\bibitem{} Metcalfe, N., Shanks, T., Fong, R., et al, 1995, MNRAS 273, 257
\bibitem{} Metcalfe, L., Altieri, B., McBreen, B., et al, 1999, in
'The universe as seen by ISO', eds P.Cox and M.F.Kessler, UNESCO,
Paris, ESA SP-427, p. 1019,
astro-ph/9901147
\bibitem{} Papoular, R., 1999, A$\&$A 346, 219
\bibitem{} Pozzetti, L., Madau, P., Zamorani, G., Ferguson, H.C.,
Bruzual, G.A., 1998, MNRAS, 298, 1133
\bibitem{} Puget, J.L., Abergel, A., Bernard, J.P., et al, 1996, A\&A 308, L5
\bibitem{} Puget, J.L., Lagache, G., Clements, D.L., et al, 1999, A\&A 345, 29
\bibitem{} Richards, E.A., Kellerman, K.L., Fomalont, E.B., Windhorst,
R.A., Partridge, R.B., 1998, AJ 116, 1039
\bibitem{} Rowan-Robinson, M., Oliver, S., Efstathiou, A., et al,
1999, in 'The universe as seen by ISO', eds P.Cox and M.F.Kessler,
UNESCO, Paris, ESA SP-427, astro-ph/9906273
\bibitem{} Rush, B., Malkan, M.A., Spinoglio, L., 1993, ApJS 89, 1
\bibitem{} Sanders, D.B., Mirabel, I.F., 1996, {\it Ann. Rev. Astron. Astrophys.} 34, 749
\bibitem{} Scott, D., Lagache, G., Borys, C., et al, 1999, submitted to A\&A, astro-ph/9910428
\bibitem{} Siebenmorgen, R., Abergel, A., Altieri, A., et al, 1996, A\&A 315, L69
\bibitem{} Soifer B.T., Neugebauer G., 1991, AJ 101, 354
\bibitem{} Taniguchi, Y., et al, 1997a, 'Taking ISO to the Limits', Laureijs R. \& Levine D. (ESA)
\bibitem{} Taniguchi, Y., Cowie, L.L., Sato, Y., Sanders, D., Kawara, K., 1997b, A\&A 328, L9
\bibitem{} Windhorst, R.A., Fomalont, E.B., Partridge, R.B., Lowenthal, J.D., 1993, ApJ 405, 498
\end{moriondbib}
\end{document}